\documentclass[10pt]{iopart}
\usepackage{braket}
\usepackage{graphicx}
\usepackage{iopams}
\usepackage{url}

\begin{document}
\title{Advancing numerics for the Casimir effect to experimentally relevant aspect ratios}
\author{Michael Hartmann$^1$, Gert-Ludwig Ingold$^1$ and Paulo A. Maia Neto$^2$}
\address{$^1$ Institut f\"ur Physik, Universit\"at Augsburg, 86135 Augsburg,
         Germany}
\address{$^2$ Instituto de F{\'i}sica, Universidade Federal do Rio de Janeiro,
         CP 68528, Rio de Janeiro RJ 21941-909, Brazil}
\eads{\mailto{michael.hartmann@physik.uni-augsburg.de},
      \mailto{gert.ingold@physik.uni-augsburg.de},
      \mailto{pamn@if.ufrj.br}}

\begin{abstract}
 Within the scattering theoretical approach, the Casimir force is obtained
 numerically by an evaluation of the round trip of an electromagnetic wave
 between the objects involved.  Recently [Hartmann M \etal 2017, {\it Phys.
 Rev. Lett.} {\bf 119} 043901] it was shown that a symmetrization of the
 scattering operator provides significant advantages for the numerical
 evaluation of the Casimir force in the experimentally relevant sphere-plane
 geometry. Here, we discuss in more detail how the symmetrization modifies
 the scattering matrix in the multipole basis and how computational time is
 reduced. As an application, we discuss how the Casimir force in the
 sphere-plane geometry deviates from the proximity force approximation as
 a function of the geometric parameters.
\end{abstract}

\submitto{\PS}


\maketitle
\ioptwocol

\section{Introduction}
When Casimir first derived a force between two objects induced by quantum
fluctuations of the electromagnetic vacuum \cite{Casimir1948}, he considered a
setup consisting of two infinitely extended, perfectly reflecting parallel
plates and obtained a universal expression for the pressure pushing the plates
towards each other. In a finite-size system, the Casimir force can thus be
increased by increasing the size of the plates.

\begin{figure}
 \begin{center}
  \includegraphics[width=0.5\columnwidth]{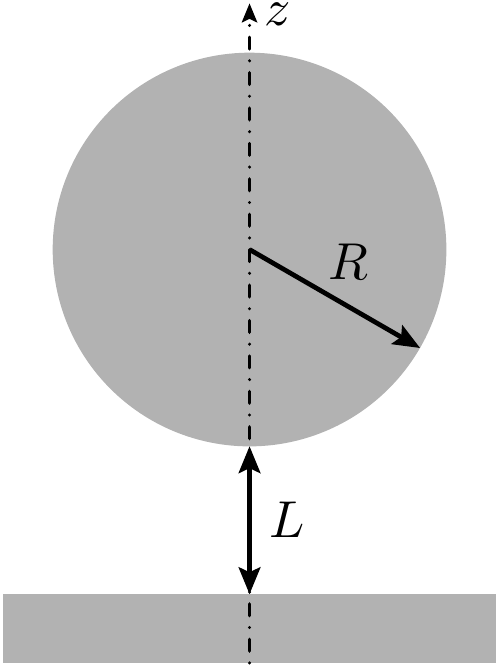}
 \end{center}
 \caption{Geometry of the sphere-plane setup. The aspect ratio $R/L$ depends
	on the sphere radius $R$ and the smallest distance $L$ between
	sphere and plane. The setup is axially symmetric about the $z$-axis.}
 \label{fig:geometry}
\end{figure}

Instead of the plane-plane geometry, most modern experiments rather employ the
sphere-plane geometry, thus avoiding the problem of misalignment. In experiments
with a particularly large sphere radius, only a section of a sphere is used. The
geometrical parameter characterizing the sphere-plane setup displayed in
\fref{fig:geometry} is the aspect ratio $R/L$, i.e.\ the ratio between the sphere
radius $R$ and the minimal distance $L$ between sphere and plane. Note that in
\fref{fig:geometry} the sphere radius is chosen to be rather small. The
corresponding aspect ratio of $4/3$ is more than two orders of magnitude smaller
than common aspect ratios used in experiments. An intuitive idea of a typical
aspect ratio is obtained by imagining the ratio of the earth radius and the
typical cruising altitude of a commercial jet.

From an experimental point of view, the aspect ratio should be chosen large in
order to obtain a sizeable force and to allow for precise measurements which
are relevant in various respects. Precise Casimir force measurements are
crucial to detect possible deviations from the gravitational interaction at
submicrometer distances \cite{Decca2003a,Decca2007} and thus to exclude or
possibly support proposed mechanisms for a fifth fundamental interaction
\cite{Fischbach1992,Antoniadis2011}. Precise measurements render the results
also sensitive to aspects of the experimental setup beyond the geometrical
features, notably the material properties of sphere and plate
\cite{Klimchitskaya2009}. This has led to the so-called Drude-plasma
controversy. While the materials used in experiments clearly have a finite
dc~conductivity which can be accounted for within the Drude model, the plasma model
with its infinite dc~conductivity describes better most Casimir force
measurements (for a recent experiment addressing this issue see
Ref.~\cite{Bimonte2016}).

While on the experimental side, one aims for relatively large aspect ratios,
on the theoretical side, the situation is a bit more complicated. The cases of
extremely large aspect ratios and of very small aspect ratios can be treated
rather easily while intermediate aspect ratios can be very demanding to
cope with.

For the theoretical description of experiments, frequently the proximity force
approximation (PFA) has been employed. It assumes that the Casimir force
between a sphere and a plate can be decomposed into contributions from many
small plane-plane segments which are integrated over. As the Casimir force is
not additive, this approach can only be approximate. For perfect
reflectors at zero temperature, the PFA yields the Casimir force in the
sphere-plane geometry
\begin{equation}
 F_\mathrm{PFA} = -\frac{\pi^3\hbar c}{360}\frac{R}{L^3}\,.
\end{equation}
The proximity force approximation can be generalized to account for arbitrary
electromagnetic response of sphere and plate as well as arbitrary temperatures
by using the corresponding expression for the force in the plane-plane geometry.
Recently, it has been proven for the sphere-sphere geometry that the
proximity force approximation yields the leading small-distance behaviour for
arbitrary temperatures and materials \cite{Spreng2018}. In particular, the
proximity force approximation for the sphere-plane geometry is obtained by
taking one sphere radius to infinity. Furthermore, Ref.~\cite{Spreng2018} has
established a connection between the specular-reflection limit of Mie scattering
and the proximity force approximation.

In the opposite limit of small aspect ratios, it is convenient to express the
Casimir force in terms of a multipole expansion. For very small aspect ratios,
the dipole approximation allows for analytical results which have been used to
analyse the appearance of a negative Casimir entropy
\cite{Ingold2015,Umrath2015}. The number of multipole moments required in a
numerical calculation of the Casimir force increases linearly with the aspect
ratio. Since typically the numerical effort grows with the third power of the
dimension, the numerical evaluation of the Casimir force has been restricted to
aspect ratios $R/L \lesssim 100$ \cite{CanaguierDurand2011}.

\begin{figure}
 \begin{center}
  \includegraphics[width=\columnwidth]{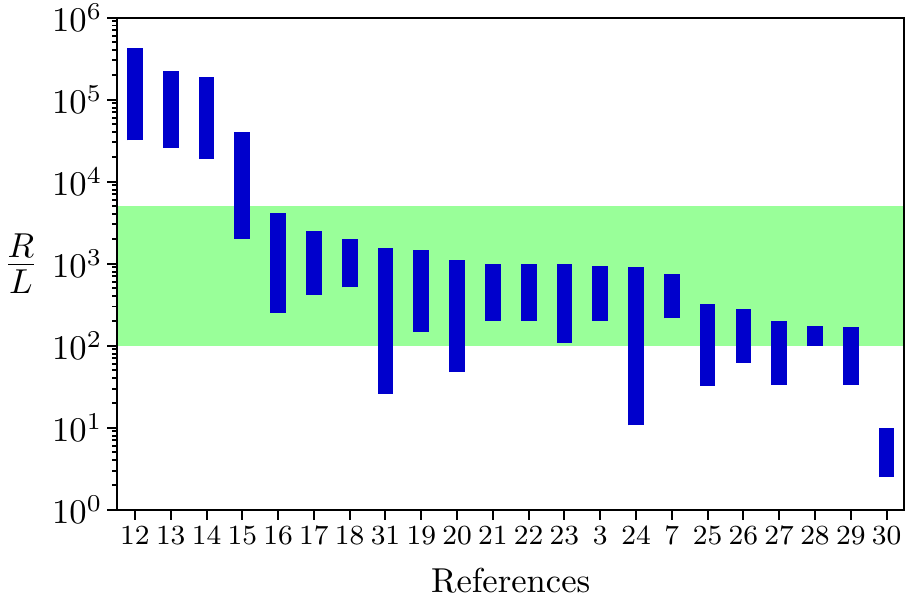}
 \end{center}
 \caption{The aspect ratio $R/L$ is shown for Casimir experiments involving
	a sphere or a spherical lens and a plane or another sphere. The range
	of aspect ratios, which became accessible through the numerical approach
	discussed here, is marked in green.}
 \label{fig:experiments}
\end{figure}

In \fref{fig:experiments}, we show the aspect ratio $R/L$ for recent
experiments
\cite{Decca2007,Bimonte2016,Masuda2009,Sushkov2011,Lamoreaux1997,GarciaSanchez2012,%
vanZwol2008,Zwol2008,deMan2009,Decca2003,Chan2001,Munday2008,Munday2009,Mohideen1998,%
Krause2007,Elzbieciak-Wodka2014,Banishev2013,Jourdan2009,Chang2012,%
Torricelli2011} measuring the Casimir force between a sphere and a plate. For
large aspect ratios, the sphere radius is typically so large that the sphere is
replaced by a sphere segment. In two of the experiments, measurements were
performed on the sphere-sphere geometry \cite{Ether2015,Garrett2018}. The
smallest aspect ratio is reached by an experiment proposing extremely sensitive
measurements of the Casimir force between two spheres by means of optical
tweezers \cite{Ether2015}.

\Fref{fig:experiments} indicates that most experiments were out of reach of
exact theoretical calculations until very recently. In order to cover a large
number of them, it is necessary to numerically treat aspect ratios $R/L\lesssim
5000$. The extension to the aspect ratios marked in green in
\fref{fig:experiments} became possible by symmetrizing the round-trip operator
describing the scattering of electromagnetic waves between sphere and plate
\cite{Hartmann2017}. The very large aspect ratios not yet covered are rather
well described by the proximity force approximation up to corrections typically
smaller than one percent.

In the following, we discuss how it becomes possible to advance into the green
region of \fref{fig:experiments}. We start in \Sref{sec:symmetrization} by
explaining as the central idea the symmetrization of the round-trip operator.
Sections~\ref{sec:operators} and \ref{sec:matrix_elements} provide technical
details allowing to set up the required matrix elements. Numerical aspects related to
hierarchical matrices and to the performance achieved by making use of them
are given in \Sref{sec:hodlr}. As a physical application, we discuss in
\Sref{sec:corrections} the size of the corrections to PFA as a function of the
geometrical parameters in the sphere-plane geometry before presenting our
conclusions in \Sref{sec:conclusions}. Some more technical details are given in
the appendices.

\section{Symmetrization of the round-trip operator}
\label{sec:symmetrization}

The Casimir force for the setup originally considered by Casimir, i.e.\ two parallel
perfectly reflecting plates, can be evaluated by summing the vacuum energy over
all modes of the electromagnetic field \cite{Schleich2001}. For more general situations
like the experimentally relevant sphere-plane geometry discussed here, the scattering
approach has turned out to be very well suited \cite{IngoldLambrecht2015}.

\begin{figure}
 \begin{center}
  \includegraphics[width=0.8\columnwidth]{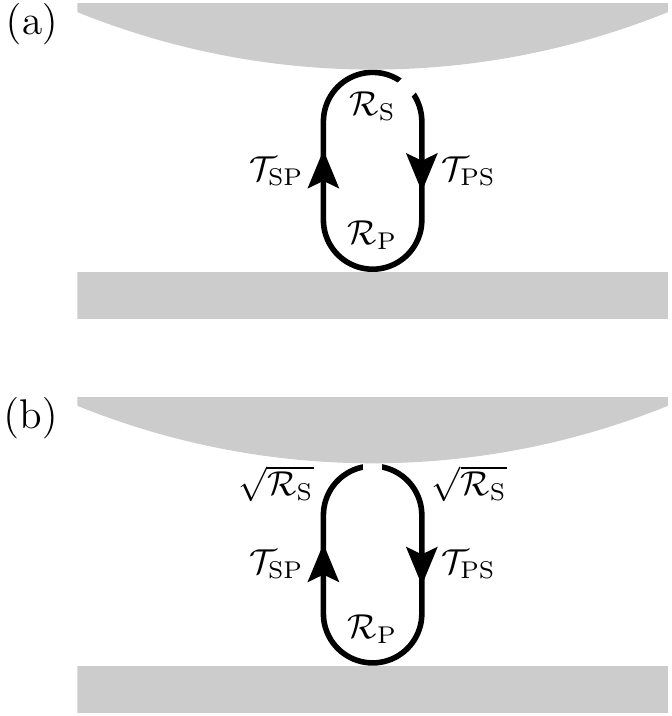}
 \end{center}
 \caption{Graphical representation of the round-trip operator between sphere
	and plane in (a) its usual form (\ref{eq:M}) and (b) the symmetrized
	form (\ref{eq:M_symm}). The reflection operators at plane and sphere
	are denoted by $\mathcal{R}_\mathrm{P}$ and $\mathcal{R}_\mathrm{S}$,
	respectively, while $\mathcal{T}_\mathrm{PS}$ and $\mathcal{T}_\mathrm{SP}$
	denote the translation operators from sphere to plane and vice versa.}
 \label{fig:roundtrip}
\end{figure}

Within the scattering approach to the Casimir effect in imaginary frequencies
$\omega=\rmi\xi$, the free energy is expressed as a sum
\cite{Lambrecht2006,Emig2007}
\begin{equation}
\mathcal{F} = \frac{k_B T}{2} \sum_{n=-\infty}^\infty \log\det\left[1-\mathcal{M}(|\xi_n|)\right]
\label{eq:matsubaraSum}
\end{equation}
over the Matsubara frequencies $\xi_n=2\pi n k_B T/\hbar$. The round-trip operator
\begin{equation}
\label{eq:M}
\mathcal{M} = \mathcal{R}_\mathrm{S} \mathcal{T}_\mathrm{SP} \mathcal{R}_\mathrm{P} \mathcal{T}_\mathrm{PS}
\end{equation}
represents a complete round-trip of an electromagnetic wave between the sphere
and the plane as indicated in \fref{fig:roundtrip}a. The operator
$\mathcal{T}_\mathrm{SP}$ describes a translation from the reference frame of
the plane to that of the sphere, and vice versa for $\mathcal{T}_\mathrm{PS}$.
$\mathcal{R}_\mathrm{S}$ denotes the reflection at the sphere, while
$\mathcal{R}_\mathrm{P}$ denotes the reflection at the plane. The Matsubara sum
\eref{eq:matsubaraSum} together with \eref{eq:M} holds even if sphere, plane or
the medium in between are dissipative \cite{Guerout2018}.

We express the round-trip operator $\mathcal{M}$ in the multipole basis
${\ket{\ell,m,P}}$ described by the angular momentum quantum numbers $\ell$ and $m$ ($\ell=1,2,\dots$ and $|m| \le \ell$). The
polarization represents electric multipoles for $P=\mathrm{E}$ and magnetic
multipoles for $P=\mathrm{M}$, respectively. Due to the rotational symmetry of
the plane-sphere setup about the $z$-axis, the round-trip operator is diagonal
in $m$ and every block $\mathcal{M}^{(m)}$ yields an independent contribution to
the free energy
\begin{equation}
\mathcal{F} = \frac{k_B T}{2} \sum_{n=-\infty}^\infty \sum_{m=-\infty}^\infty \log\det\left(1-\mathcal{M}^{(m)}(|\xi_n|)\right)\,.
\label{eq:matsubaraSumM}
\end{equation}

\begin{figure}
 \begin{center}
  \includegraphics[width=\columnwidth]{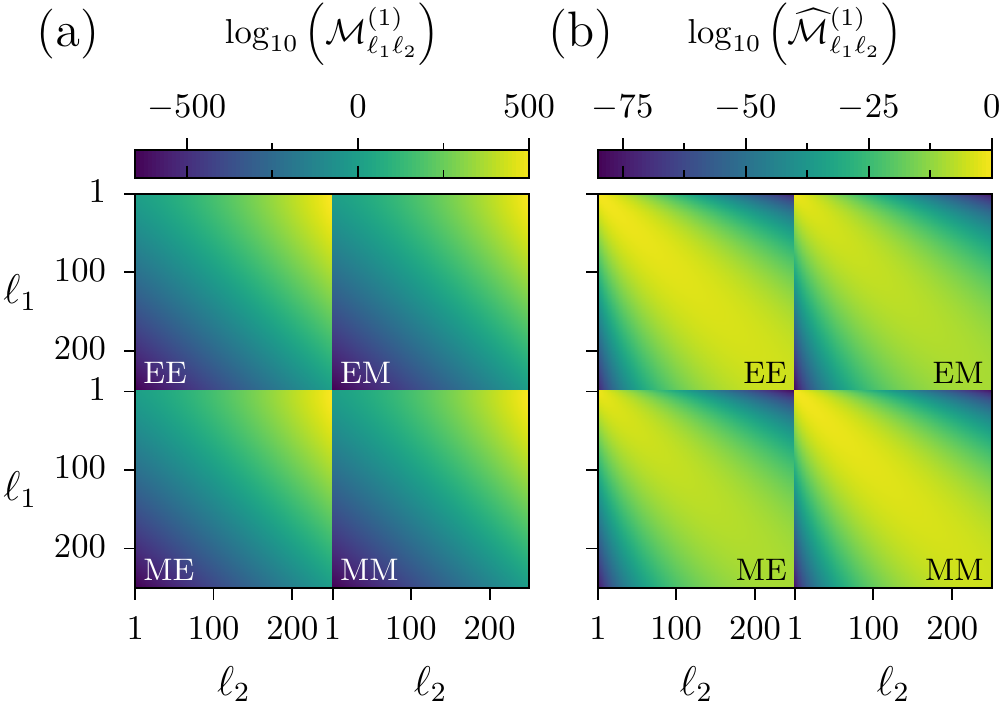}
 \end{center}
 \caption{The logarithm of the matrix elements of (a) the non-symmetrized
   round-trip operator $\mathcal{M}^{(m)}$ and (b) the symmetrized round-trip
   operator $\widehat{\mathcal{M}}^{(m)}$ in the multipole basis is shown on
   a colour scale for $R/L=50$, $\xi(L+R)/c=1$, $m=1$, and perfect reflectors.
   The four blocks correspond to the different sequences of polarizations during
   a round trip. While the non-symmetrized round-trip matrix is ill-conditioned,
   the matrix elements of the symmetrized round-trip operator take their maximum on
   the diagonal and decrease away from it.}
 \label{fig:matrix_elems}
\end{figure}

The numerical problems associated with the definition \eref{eq:M} of the
round-trip operator become clear from \fref{fig:matrix_elems}a where we depict
the values of the matrix elements $\langle \ell_1, m, P_1\vert\mathcal{M}
\vert\ell_2, m, P_2\rangle$ on a logarithmic colour scale. Even though the
matrix elements depend on the parameters chosen here, $R/L=50$, $\xi(L+R)/c=1$,
$m=1$, and perfect conductors, the data shown are typical. Already for the
relatively small aspect ratio used in \fref{fig:matrix_elems}a, the round-trip
operator \eref{eq:M} clearly results in an ill-conditioned matrix with elements
differing by hundreds of orders of magnitude. As a consequence, a fast and stable
numerical evaluation of the determinant becomes extremely difficult. When the
determinants in \eref{eq:matsubaraSumM} are evaluated, the combination of very
small and very large matrix elements can yield contributions of the order one.
Small perturbations in the matrix elements may then cause large errors.
Furthermore, common computer number formats cover a range of numbers from about
$10^{-324}$ to $10^{308}$ \cite{IEEE754} which is not sufficient to represent
all matrix elements in \fref{fig:matrix_elems}a. Instead, one has to use number
formats that cover a wider range of numbers, but are also significantly slower.

To overcome these problems, we make use of the fact that the round-trip
operator is not uniquely defined. Instead of \eref{eq:M}, we choose the
symmetrized form for the round-trip operator
\begin{equation}
\label{eq:M_symm}
\widehat{\mathcal{M}} = \sqrt{\mathcal{R}_\mathrm{S}} \mathcal{T}_\mathrm{SP}
    \mathcal{R}_\mathrm{P} \mathcal{T}_\mathrm{PS} \sqrt{\mathcal{R}_\mathrm{S}} \,.
\end{equation}
as illustrated in \fref{fig:roundtrip}b. The scattering operator at the
sphere $\mathcal{R}_\mathrm{S}$ is diagonal in the multipole basis and
therefore the matrix square root exists. With this choice for the round-trip
operator, the matrix elements take their maximum on the diagonal of each
polarization block and decrease away from it, as can be seen in
\fref{fig:matrix_elems}b. Here, the same parameters have been used as in
\fref{fig:matrix_elems}a.

\begin{figure}
 \begin{center}
  \includegraphics[width=\columnwidth]{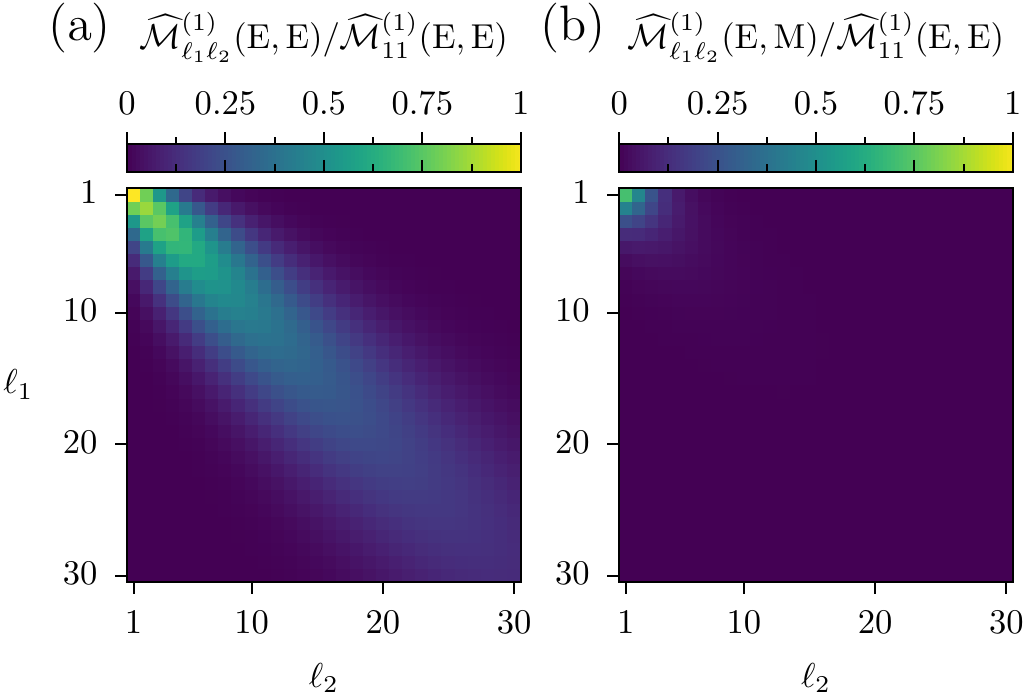}
 \end{center}
 \caption{Details of the matrix elements for (a) the polarization maintaining
 block $(\mathrm{E}, \mathrm{E})$ and (b) the polarization mixing block
 $(\mathrm{E}, \mathrm{M})$ are shown for the same parameters as in
 \fref{fig:matrix_elems}, but now on a linear scale. All matrix elements are
 taken with respect to the largest matrix element in the round-trip matrix.}
 \label{fig:matrix_elems_lin}
\end{figure}

In \fref{fig:matrix_elems_lin}, we depict a polarization-conserving block
(left) and a polarization-mixing block (right) with the matrix elements taken
from \fref{fig:matrix_elems}b, but now on a linear colour scale. This
representation emphasizes the fact that a sizeable fraction of the matrix
elements off the diagonal is numerically irrelevant. Furthermore, the contribution
of the polarization-mixing blocks is significantly smaller than that of the
polarization-conserving blocks.

\section{Reflection and translation operators}
\label{sec:operators}

The reflection operator at the sphere is diagonal in the multipole basis
\begin{equation}
\label{eq:Rs}
\mathcal{R}_\mathrm{S}\vert\ell, m, P\rangle =
    r_{\ell,P}^\mathrm{(S)}\vert\ell, m, P\rangle\,.
\end{equation}
The reflection coefficients
\begin{equation}
r_{\ell, E}^\mathrm{(S)} = -a_\ell, \quad r_{\ell, M}^\mathrm{(S)} = -b_\ell
\end{equation}
are given by the Mie coefficients $a_\ell$ and $b_\ell$, where the minus sign
is a consequence of the definition of the Mie coefficients employed here
\cite{BohrenHuffman}. For explicit expressions, we refer the reader to
\ref{appendix:mie}.

While the multipole basis is well adapted for the scattering at the sphere,
plane waves are better suited to describe the reflection at the plate and the
translation between the reference frames of plate and sphere.
A plane wave is characterized by the wave vector $\mathbf{K}$ and
the polarization $p$. Given the special role of the $z$-axis as symmetry axis,
we choose the basis vectors for transverse electric (TE) and transverse magnetic
(TM) modes as
\begin{equation}
\hat{\varepsilon}_\mathrm{TE} = \frac{\hat{\mathbf{z}}\times\hat{\mathbf{K}}}{|\hat{\mathbf{z}}\times\hat{\mathbf{K}}|}, \quad
\hat{\varepsilon}_\mathrm{TM} = \hat{\varepsilon}_\mathrm{TE}\times\hat{\mathbf{K}} \,.
\end{equation}
Here, unit vectors are denoted by a hat. Within the Matsubara formalism, all
quantities are expressed in terms of imaginary frequencies $\xi$. For the wave
vector $\mathbf{K}$, we choose an imaginary $z$-component $\kappa$ while the
projection $\mathbf{k}$ onto the $x$-$y$-plane is kept real. The dispersion
relation then reads
\begin{equation}
\xi^2 = c^2(\kappa^2-\vert\mathbf{k}\vert^2)
\label{eq:dispersionsrelation}
\end{equation}
with the speed of light $c$. As the frequency remains constant during a round
trip, it is convenient to make use of the angular spectral representation
\cite{Nieto-Vesperinas2006}. The corresponding basis
$\{\ket{\mathbf{k},p,\phi}\}$ is then labeled by the projection of the
wave-vector $\mathbf{K}$ onto the $x$-$y$-plane, $\mathbf{k}=(k_x,k_y,0)$, the
polarization $p=\mathrm{TE},\mathrm{TM}$, and the direction of propagation
$\phi=\pm1$ in $\pm z$-direction. The latter fixes the sign ambiguity
when solving \eref{eq:dispersionsrelation} for $\kappa$. We are
therefore free to always choose $\kappa$ as positive.

Plane waves propagating in $-z$-direction are reflected by the plate
\begin{equation}
\mathcal{R}_\mathrm{P}\ket{\mathbf{k}, p,-} = r_p^\mathrm{(P)}(\mathrm{i}\xi,k)\ket{\mathbf{k}, p, +}
\end{equation}
where the reflection coefficients $r_p^{(\mathrm{P})}$ are the Fresnel coefficients.
Explicit expressions are given in \ref{appendix:fresnel}.

Finally, the translation operators are diagonal in the plane-wave basis
\begin{eqnarray}
\mathcal{T}_\mathrm{SP} \ket{\mathbf{k},p,+} &= \rme^{-\kappa(L+R)} \ket{\mathbf{k},p,+}, \\
\mathcal{T}_\mathrm{PS} \ket{\mathbf{k},p,-} &= \rme^{-\kappa(L+R)} \ket{\mathbf{k},p,-}
\end{eqnarray}
with matrix elements given by exponential factors. Here, we have assumed
vacuum between sphere and plane as we will do in the rest of the paper.

\section{Matrix elements of the round-trip operator}
\label{sec:matrix_elements}

Having discussed the individual building blocks in the previous section,
we now combine them to obtain the matrix elements of the symmetrized
round-trip matrix \eref{eq:M_symm} in the multipole basis
\begin{eqnarray}
\label{eq:Melem}
\widehat{\mathcal{M}}_{\ell_1\ell_2}^{(m)}(P_1,P_2)
=\braket{\ell_1,m,P_1|\widehat{\mathcal{M}}|\ell_2,m,P_2}\nonumber\\
= \sqrt{r_{\ell_1,P_1}^\mathrm{(S)}} \sqrt{r_{\ell_2,P_2}^\mathrm{(S)}}
    \sum_p \int_0^\infty \frac{\rmd^2\mathbf{k}}{(2\pi)^2} r_p^\mathrm{(P)} \rme^{-2\kappa(L+R)}\\
\qquad\times \braket{\ell_1,m,P_1|\mathbf{k},p,+}\nonumber
    \braket{\mathbf{k},p,- | \ell_2,m,P_2} \,.\nonumber
\end{eqnarray}
The last two factors arise due to a change from the multipole basis to the
plane-wave basis and vice versa between the translation operators and the
reflection operators at the sphere. Explicit expressions for these matrix
elements are given in Ref.~\cite{Messina2015}.

We organize the round-trip matrix in the form of a block matrix
\begin{equation}
\widehat{\mathcal{M}}^{(m)} = \left(
\begin{array}{cc}
    \widehat{\mathcal{M}}^{(m)}(\mathrm{E},\mathrm{E}) &
    \widehat{\mathcal{M}}^{(m)}(\mathrm{E},\mathrm{M}) \\
    \widehat{\mathcal{M}}^{(m)}(\mathrm{M},\mathrm{E}) &
    \widehat{\mathcal{M}}^{(m)}(\mathrm{M},\mathrm{M})
\end{array} \right)
\end{equation}
where the diagonal blocks correspond to matrix elements preserving polarization,
and the off-diagonal blocks correspond to matrix elements with a change of
polarization. Expressing the double integral in \eref{eq:Melem} in polar
coordinates, the integral over the angle can be carried out. After a similarity
transform of the round-trip matrix in order to remove phase factors, we finally
obtain the matrix elements
\begin{eqnarray}
\widehat{\mathcal{M}}_{\ell_1\ell_2}^{(m)}(\mathrm{M},\mathrm{M})
 &= \sqrt{|b_{\ell_1} b_{\ell_2}|} \left( A_{\ell_1\ell_2,\mathrm{TM}}^{(m)} - B_{\ell_1\ell_2,\mathrm{TE}}^{(m)} \right)
\label{eq:matrix_elems_sym_MM}\\
\widehat{\mathcal{M}}_{\ell_1\ell_2}^{(m)}(\mathrm{E},\mathrm{E})
 &= \sqrt{|a_{\ell_1} a_{\ell_2}|} \left( B_{\ell_1\ell_2,\mathrm{TM}}^{(m)} - A_{\ell_1\ell_2,\mathrm{TE}}^{(m)} \right)
\label{eq:matrix_elems_sym_EE}\\
\widehat{\mathcal{M}}_{\ell_1\ell_2}^{(m)}(\mathrm{E},\mathrm{M})
 &= \sqrt{|a_{\ell_1} b_{\ell_2}|} \left( C_{\ell_1\ell_2,\mathrm{TM}}^{(m)} - C_{\ell_2\ell_1,\mathrm{TE}}^{(m)} \right)
\label{eq:matrix_elems_sym_EM}\\
\widehat{\mathcal{M}}_{\ell_1\ell_2}^{(m)}(\mathrm{M},\mathrm{E})
 &= \widehat{\mathcal{M}}_{\ell_2\ell_1}^{(m)}(\mathrm{E},\mathrm{M})\,,
\label{eq:matrix_elems_sym_ME}
\end{eqnarray}
which depend on the integrals
\begin{eqnarray}
A_{\ell_1\ell_2,p}^{(m)} &= \int_1^\infty \rmd x f^{(m)}_{\ell_1,\ell_2,p}(x, -1)P_{\ell_1}^m(x) P_{\ell_2}^m(x) \\
B_{\ell_1\ell_2,p}^{(m)} &= \int_1^\infty \rmd x f^{(m)}_{\ell_1,\ell_2,p}(x, 1){P_{\ell_1}^m}^{\prime}(x){P_{\ell_2}^m}^{\prime}(x) \\
C_{\ell_1\ell_2,p}^{(m)} &= \int_1^\infty \rmd x f^{(m)}_{\ell_1,\ell_2,p}(x, 0)P_{\ell_1}^m(x){P_{\ell_2}^m}^{\prime}(x)
\end{eqnarray}
where
\begin{eqnarray}
f^{(m)}_{\ell_1,\ell_2,p}(x, j) &= m\Lambda^{(m)}_{\ell_1}\Lambda^{(m)}_{\ell_2}
    r_p^{(\mathrm{P})}\big(\mathrm{i}\xi, \frac{\xi}{c}\sqrt{x^2-1}\big) \nonumber \\
&\quad\times\left(\frac{x^2-1}{m}\right)^j\exp\left(-2\frac{\xi(L+R)}{c}x\right)
\end{eqnarray}
and
\begin{equation}
\Lambda_{\ell}^{(m)} = \sqrt{\frac{2\ell+1}{\ell(\ell+1)}\frac{(\ell-m)!}{(\ell+m)!}} \,.
\end{equation}
Note that for the associated Legendre polynomials $P_{\ell}^m$, we use
an uncommon phase convention defined in \ref{appendix:plm}. The dimension of
the matrices (\ref{eq:matrix_elems_sym_MM})--(\ref{eq:matrix_elems_sym_ME}) is
infinite. For a numerical evaluation, the vector space has to be truncated in
the angular momentum.

The matrix elements for the round-trip operator $\mathcal{M}$ differ from those
of the symmetrized round-trip operator $\widehat{\mathcal{M}}$ only with respect
to the Mie coefficients. While for the symmetrized round-trip operator according
to \eref{eq:matrix_elems_sym_MM}--\eref{eq:matrix_elems_sym_ME} the matrix
elements are proportional to the square root of a product of Mie coefficients
with different angular momenta, the matrix elements of $\mathcal{M}$ are
proportional to one Mie coefficient with angular momentum $\ell_1$, thus
resulting in the numerical problems discussed above.

As the Fresnel coefficients are positive for $p=\mathrm{TM}$ and negative for
$p=\mathrm{TE}$, the matrix elements of the round-trip operator
$\widehat{\mathcal{M}}^{(m)}$ are positive. Also, as the integrals
$A_{\ell_1\ell_2,p}^{(m)}$ and $B_{\ell_1\ell_2,p}^{(m)}$ are symmetric with
respect to $\ell_1$ and $\ell_2$, the round-trip matrix is symmetric. Numerical
tests suggest that the scattering matrix $1-\widehat{\mathcal{M}}^{(m)}$ is
diagonally dominant. Together with the positivity of the diagonal entries, it
follows that the scattering matrix $1-\widehat{\mathcal{M}}^{(m)}$ is positive
definite. These properties ensure the stability of the numerical evaluation of
the determinant. Small perturbations in the matrix elements thus only cause
small changes in the value of the determinant \cite{Dailey2014}.

\section{Hierarchical matrices}
\label{sec:hodlr}

The symmetrization allows to exploit further properties of the round-trip
matrices. As the matrix elements decrease away from the diagonal, one might
think that the dominant contribution to the determinant comes from matrix
elements close to the diagonal. In fact, it turns out that the matrices
$\mathcal{M}^{(m)}$ are hierarchical off-diagonal low-rank (HODLR) matrices
\cite{Ambikasaran2013}. This means that the round-trip matrices can be
sub-divided into a hierarchy of rectangular blocks which can be approximated by
low-rank matrices.

A low-rank matrix $M$ of dimension $N\times N$ can be efficiently approximated
by
\begin{equation}
\label{eq:lowrank}
M \approx U V^T
\end{equation}
where $U$ and $V$ are matrices of dimension $N\times p$ with $p \ll N$. The
best rank $p$ approximation of $M$ can be obtained using the singular value
decomposition \cite{Eckart1936}. Instead of a computationally expensive full
singular-value decomposition, low-rank approximations can be computed using fast
algorithms like adaptive cross approximation, pseudo-skeletal approximations,
interpolatory decompositions or rank revealing QR and LU (see
\cite{Ambikasaran2013} and references therein).

\begin{figure}
 \begin{center}
  \includegraphics[width=\columnwidth]{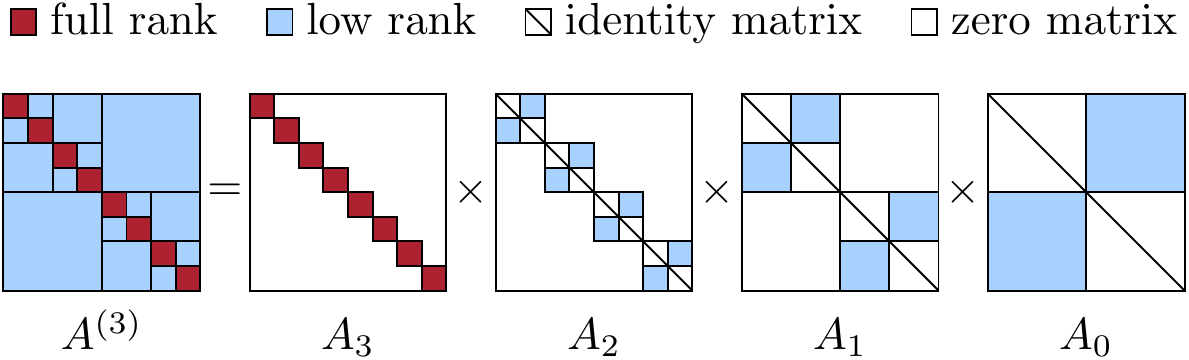}
 \end{center}
    \caption{Graphical representation of the factorization
    \eref{eq:factorization} of a HODLR matrix for $n=3$. Red blocks represent
    full-rank matrices while bright blue blocks correspond to low-rank
    matrices (after \cite{Ambikasaran2013}).}
 \label{fig:hodlr}
\end{figure}

A HODLR matrix $A$ can be factored into $n+1$ block-diagonal matrices
\begin{equation}
\label{eq:factorization}
A \approx A^{(n)} = A_n A_{n-1} \dots A_0
\end{equation}
as sketched in \fref{fig:hodlr} for $n=3$. The matrix $A_n$ consists of $2^n$
full-rank blocks around the diagonal while the other matrices $A_{n-1}$ to $A_0$
represent low-rank updates to the identity. The error of this approximation can
be made negligibly small by choosing appropriate ranks $p$.

The factorization (\ref{eq:factorization}) allows for a fast computation of the
determinant $A^{(n)}$ for two reasons. Firstly, one can exploit the
multiplicativity of determinants. Secondly, the block matrices appearing in
$A_0$ to $A_{n-1}$ are of the form
\begin{equation}
\label{eq:matrix_b}
B = \left(\begin{array}{cc}
1 & U_1V_1^T \\
U_2V_2^T & 1
\end{array}\right)\,,
\end{equation}
requiring at first sight the evaluation of the determinant of an
$N\times N$-matrix according to
\begin{equation}
\label{eq:det1}
 \det(B) = \det\left(1 - U_1 V_1^T U_2 V_2^T\right) \,.
\end{equation}
However, exploiting Sylvester's determinant identity
\begin{equation}
\det(1+AB) = \det(1+BA)\,,
\end{equation}
we obtain
\begin{equation}
\label{eq:det2}
 \det(B) = \det\left(1 - V_2^T U_1 V_1^T U_2\right) \,.
\end{equation}
It is thus sufficient to evaluate the determinant of a $p\times p$-matrix,
resulting in a significant speed-up.

In order to assess the numerical advantages of the approach discussed above, we
compute the determinants of the scattering matrices either by means of a
Cholesky decomposition or the implementation \cite{hodlrcode} of the algorithm
for HODLR matrices described in \cite{Ambikasaran2013}. The Cholesky
decomposition factorizes a symmetric positive-definite matrix into the product
of a triangular matrix and its transpose allowing for a simple computation of
the determinant. The factorization requires $\mathcal{O}(N^3)$ of time for an
$N\times N$ matrix and is about twice as fast as an LU decomposition. The
computation of determinants using the HODLR approach takes $\mathcal{O}(p^2
N\log^2 N)$ steps where, depending on the nature of the problem, $p$ may be a
function of $N$.

\begin{figure}
 \begin{center}
  \includegraphics[width=\columnwidth]{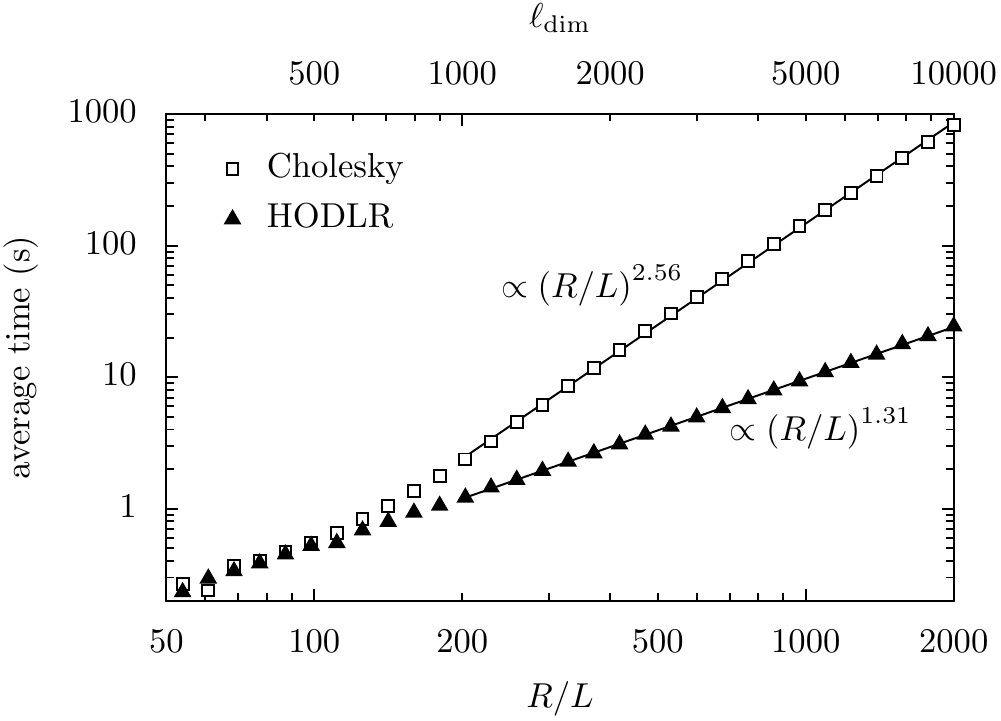}
 \end{center}
 \caption{Average runtime to compute
 $\log\det\left(1-\mathcal{M}^{(m)}(\xi)\right)$ as a function of the aspect
 ratio $R/L$ for $m=1$, $\xi=c/(L+R)$, and perfect reflectors. The angular
 momentum is truncated at $\ell_\mathrm{dim} = 5R/L$, yielding a scattering
 matrix of dimension $2\ell_\mathrm{dim}\times2\ell_\mathrm{dim}$. Squares and
 triangles correspond to a computation using Cholesky decomposition and the
 HODLR algorithm \cite{hodlrcode}, respectively. The lines represent fits and
 correspond to the asymptotic scaling of the two algorithms. The computations
 were carried out on an Intel Core i7 with 3.4\,GHz. For the Cholesky
 decomposition LAPACK \cite{lapack} in combination with ATLAS \cite{atlas} was used.}
 \label{fig:performance}
\end{figure}

In \fref{fig:performance} we compare the average time to compute
$\log\det\left(1-\mathcal{M}^{(m)}(\xi)\right)$ depending on the aspect ratio
using the HODLR approach and a Cholesky decomposition. We specifically choose
$m=1$, $\xi=c/(L+R)$, and perfect reflectors, but other parameters yield similar
results. For aspect ratios $R/L \lesssim 100$, both algorithms take about the
same time. For larger aspect ratios $R/L \gtrsim 200$, the computational time
using the Cholesky decomposition scales as $\propto (R/L)^{2.56}$. This is
faster than the theoretical complexity $\mathcal{O}((R/L)^3)$ of the Cholesky
decomposition, because our numerical implementation saves and reuses
intermediate values so that the computational time per matrix element is not
constant. In contrast, the HODLR algorithm becomes significantly faster for $R/L
\gtrsim 200$ and the computational time scales only as $\propto (R/L)^{1.31}$.
For the largest aspect ratio displayed in \fref{fig:performance}, $R/L=2000$, we
find a speed-up by a factor 33. At even larger aspect ratios, the time required
by the Cholesky decomposition becomes prohibitively long.

\section{Corrections to the PFA for the force}
\label{sec:corrections}

\begin{figure}
 \begin{center}
  \includegraphics[width=\columnwidth]{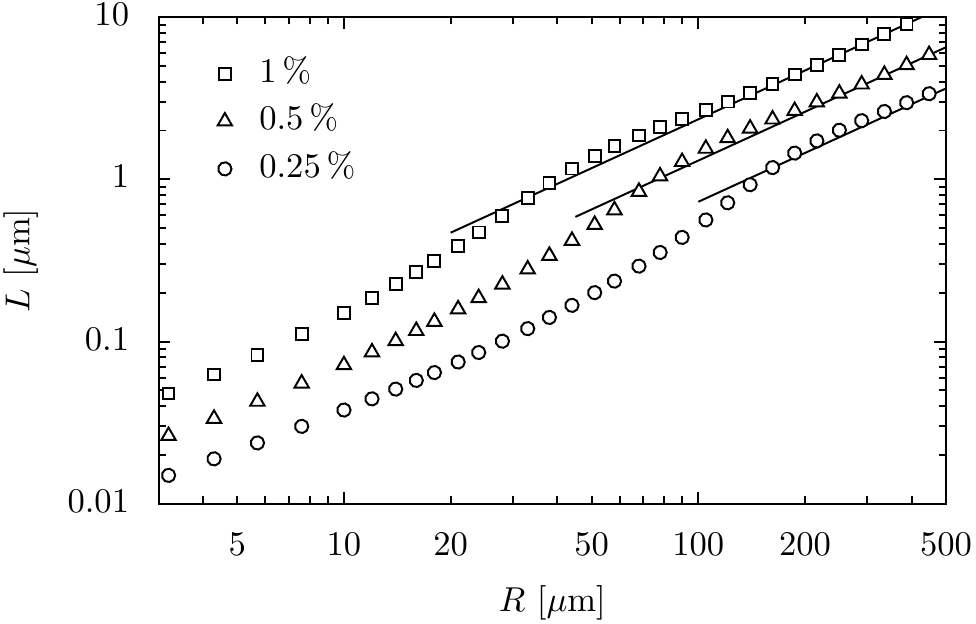}
 \end{center}
 \caption{The correction $1-F/F_\mathrm{PFA}$ of the force with respect to the
    corresponding PFA result is displayed as a function of the radius $R$
    of the sphere and its distance $L$ from the plane. The numerical data
    depicted by symbols corresponding to different values of the correction
    have been obtained for gold at $T=300\,\mathrm{K}$ as explained in the
    text. The solid lines indicate the corresponding high-temperature result
    according to Ref.~\cite{Bimonte2012}.}
 \label{fig:forcecorrection}
\end{figure}

As an application of our numerical approach, we consider the corrections to the
Casimir force
\begin{equation}
    F = -\frac{\partial \mathcal{F}}{\partial L}
\end{equation}
beyond the proximity force approximation. The geometrical
parameters of the sphere-plane geometry $R$ and $L$ (cf.\
\fref{fig:geometry}) for which the relative correction $1-F/F_\mathrm{PFA}$
of the force $F$ with respect to its PFA value $F_\mathrm{PFA}$ takes the values
$0.25\,\%, 0.5\,\%$, and $1\,\%$ can be read off from the corresponding symbols
shown in \fref{fig:forcecorrection}. The numerical data have been
determined for gold surfaces at a finite temperature of $300\,\mathrm{K}$.
According to \ref{appendix:mie} and \ref{appendix:fresnel}, we need the
dielectric function $\epsilon(\mathrm{i}\xi)$ for gold on the imaginary axis
which can be obtained from tabulated data \cite{Palik1998} by means of a
procedure explained in Ref.~\cite{Lambrecht2000}. In view of the still ongoing
debate on how to correctly treat the zero-frequency contribution to the
Matsubara sum, we note that here we use the Drude prescription where the
transverse electric mode does not contribute. For the plasma prescription,
the corresponding curves for a given value of the correction with respect
to PFA lie below the ones shown in \fref{fig:forcecorrection}, i.e.\ at larger
aspect ratios \cite{Hartmann2017}.

For a given value of the correction to the PFA force, the data in
\fref{fig:forcecorrection} display three different regimes. In the upper right
corner, i.e.\ for large values of $L$ and $R$, the thermal wavelength
$\lambda_\mathrm{T} = \hbar c/k_\mathrm{B}T$ can be taken to be small, thus
indicating the high-temperature regime. In contrast, the low-temperature
behaviour is found in the opposite corner, i.e.\ the lower left part.
The two regimes are joined by a transition region visible in the middle
of \fref{fig:forcecorrection}.

In the high-temperature regime, only the term with $n=0$ in the Matsubara sum
\eref{eq:matsubaraSum} contributes and thus only the zero-frequency limit of
the dielectric function is relevant. This special case within the Drude model
allows for an analytical solution for the free energy in the sphere-plane setup
\cite{Bimonte2012}. As $R$ and $L$ are the only length scales remaining in the
high-temperature limit, a fixed value of the correction to the PFA force yields a
straight line of slope 1 in the presentation of \fref{fig:forcecorrection}.
The high-temperature limits for the chosen values of the correction are
indicated by solid lines. As expected, our numerical data approach the
high-temperature limit for large sphere radii and large distances between sphere
and plane. We note that with decreasing sphere radius, the distance $L$ required
for a given value of the correction to the PFA force exceeds the distance
obtained within the high-temperature limit. However, decreasing the radius
further, the distance falls below the prediction of the high-temperature limit.

\section{Conclusions}
\label{sec:conclusions}

The standard approach to calculating Casimir forces and free energies within
the scattering theory in the multipole basis has been plagued with
ill-conditioned round-trip matrices resulting in various numerical problems.
We have shown that these problems can be eliminated by a symmetrization of
the round-trip operator which physically amounts to splitting the round-trip
of an electromagnetic wave between the two scattering objects in the middle
of the reflection at one of these objects. The symmetrization of the round-trip
operator allows to significantly reduce computational time and thus to perform
calculations in the experimentally relevant regime of aspect ratios in the
sphere-plane geometry. Furthermore, numerical errors become controllable.

As large aspect ratios are now numerically accessible, it has become possible to
assess the quality of the proximity force approximation by determining the
deviations from the exact result (cf.\ \fref{fig:forcecorrection}). This is
particularly relevant as the analysis of experimental results so far has relied
on the proximity force approximation. Specifically, precise numerical results
are of importance in the ongoing Drude-plasma debate. As an example, with the
numerical approach detailed here it has become possible recently
\cite{Hartmann2017} to demonstrate that the experimental bounds for the Casimir
force gradient found in Ref.~\cite{Krause2007} are violated both for the Drude
and the plasma prescription, even though the violation for the latter is found
to be significantly larger.

\section*{Acknowledgments}
This paper is dedicated to Wolfgang Schleich on the occasion of his
60$^\mathrm{th}$ birthday. Particularly during our yearly Augsburg-Ulm meetings,
we could admire his sagacious physical reasoning and draw inspiration from it.

We thank A. Canaguier-Durand, R. Gu\'erout, A. Lambrecht, and S. Reynaud for
discussions and D. Dalvit for providing numerical data for the dielectric
function of gold at imaginary frequencies. The authors acknowledge support from
CAPES and DAAD through the PROBRAL collaboration program. P. A. M. N. also
thanks CNPq and FAPERJ for partial financial support.

\appendix

\section{Associated Legendre polynomials}
\label{appendix:plm}

We define the associated Legendre polynomials as \cite{Zhang1996}
\begin{equation}
P_\ell^m(x) = (x^2-1)^{m/2} \frac{\mathrm{d}^m}{\mathrm{d}x^m} P_\ell(x)
\end{equation}
where $P_\ell(x)$ denotes the ordinary Legendre polynomial of degree $\ell$.
The phase convention employed here differs from the common choice usually made
in physics: We omit the Condon-Shortley phase, and change the sign in the first
term. With this choice the associated Legendre polynomials are real and
non-negative functions for $x\ge1$.

\section{Mie coefficients}
\label{appendix:mie}

The Mie coefficients are given by \cite{BohrenHuffman}
\begin{eqnarray}
\label{eq:mie_a}
a_\ell(\rmi x) &= (-1)^\ell     \frac{\pi}{2} \frac{n^2 s_\ell^{(\mathrm{a})}-s_\ell^{(\mathrm{b})}}
                                                   {n^2 s_\ell^{(\mathrm{c})}+s_\ell^{(\mathrm{d})}}, \\
\label{eq:mie_b}
b_\ell(\rmi x) &= (-1)^{\ell+1} \frac{\pi}{2} \frac{s_\ell^{(\mathrm{b})}-s_\ell^{(\mathrm{a})}}
                                                   {s_\ell^{(\mathrm{c})}+s_\ell^{(\mathrm{d})}},
\end{eqnarray}
where
\begin{eqnarray}
s_\ell^{(\mathrm{a})}(x) &= I_{\ell+\frac{1}{2}}(nx) \left[  xI_{\ell-\frac{1}{2}}(x)  - \ell I_{\ell+\frac{1}{2}}(x)  \right], \\
s_\ell^{(\mathrm{b})}(x) &= I_{\ell+\frac{1}{2}}(x)  \left[ nxI_{\ell-\frac{1}{2}}(nx) - \ell I_{\ell+\frac{1}{2}}(nx) \right], \\
s_\ell^{(\mathrm{c})}(x) &= I_{\ell+\frac{1}{2}}(nx) \left[  xK_{\ell-\frac{1}{2}}(x)  + \ell K_{\ell+\frac{1}{2}}(x)  \right], \\
s_\ell^{(\mathrm{d})}(x) &= K_{\ell+\frac{1}{2}}(x)  \left[ nxI_{\ell-\frac{1}{2}}(nx) - \ell I_{\ell+\frac{1}{2}}(nx) \right].
\end{eqnarray}
The Mie coefficients are evaluated at $x=\xi R/c$ and
$n=\sqrt{\epsilon(\rmi\xi)}$ denotes the refractive index. The dielectric
function $\epsilon(\rmi\xi)$ is evaluated at imaginary frequencies. The
coefficients $s_\ell^{(\mathrm{a})}$, $s_\ell^{(\mathrm{b})}$,
$s_\ell^{(\mathrm{c})}$, $s_\ell^{(\mathrm{d})}$, and the numerators in
\eref{eq:mie_a} and \eref{eq:mie_b} are positive independently of $\ell$ and $\xi$.

\section{Fresnel coefficients}
\label{appendix:fresnel}

The Fresnel coefficients are given by
\begin{eqnarray}
r_\mathrm{TE}^\mathrm{(P)}(\rmi\xi,k) &= \frac{c\kappa - \sqrt{c^2\kappa^2 + \xi^2\left[\epsilon(\rmi\xi)-1\right]}}{c\kappa + \sqrt{c^2\kappa^2 + \xi^2\left[\epsilon(\rmi\xi)-1\right]}}, \\
r_\mathrm{TM}^\mathrm{(P)}(\rmi\xi,k) &= \frac{\epsilon(\rmi\xi)c\kappa - \sqrt{c^2\kappa^2 + \xi^2\left[\epsilon(\rmi\xi)-1\right]}}{\epsilon(\rmi\xi)c\kappa + \sqrt{c^2\kappa^2 + \xi^2\left[\epsilon(\rmi\xi)-1\right]}} .
\end{eqnarray}
For perfect reflectors they simplify to $r_\mathrm{TM}^\mathrm{(P)}=1$ and
$r_\mathrm{TE}^\mathrm{(P)}=-1$.

\end{document}